\documentclass[twocolumn]{aastex62}
\usepackage{aas_macros}
\usepackage{graphicx}
\usepackage{epstopdf}
\usepackage{mathrsfs}
\usepackage{hyperref}
\usepackage{amsmath}
\usepackage{xspace}
\usepackage{natbib}
\usepackage{natbib}

\begin{document}
\title{Super-Solar Metallicity Stars in the Galactic Center Nuclear Star Cluster: Unusual Sc, V, and Y Abundances}
\author{Tuan Do}
\affiliation{UCLA Galactic Center Group, Physics and Astronomy Department,UCLA, Los Angeles, CA 90095-1547}
\author{Wolfgang Kerzendorf}
\affiliation{ESO, Garching, Germany}
\author{Quinn Konopacky}
\affiliation{University of California San Diego, San Diego, CA}
\author{Joseph M. Marcinik}
\affiliation{St. Vincent College, Latrobe, PA}
\author{Andrea Ghez}
\affiliation{UCLA Galactic Center Group, Physics and Astronomy Department,UCLA, Los Angeles, CA 90095-1547}
\author{Jessica R. Lu}
\affiliation{University of California Berkeley, Berkeley, CA}
\author{Mark R. Morris}
\affiliation{UCLA Galactic Center Group, Physics and Astronomy Department,UCLA, Los Angeles, CA 90095-1547}


\correspondingauthor{Tuan Do}
\email{tdo@astro.ucla.edu}

\begin{abstract}
We present adaptive-optics assisted near-infrared high-spectral resolution observations of late-type giants in the nuclear star cluster of the Milky Way. The metallicity and elemental abundance measurements of these stars offer us an opportunity to understand the formation and evolution of the nuclear star cluster. In addition, their proximity to the supermassive black hole ($\sim 0.5$ pc) offers a unique probe of the star formation and chemical enrichment in this extreme environment. We observed two stars identified by medium spectral-resolution observations as potentially having very high metallicities. We use spectral-template fitting with the PHOENIX grid and Bayesian inference to simultaneously constrain the overall metallicity, [M/H], alpha-element abundance [$\alpha$/Fe], effective temperature, and surface gravity of these stars. We find that one of the stars has very high metallicity ([M/H] $> 0.6$) and the other is slightly above solar metallicity. Both Galactic center stars have lines from scandium (Sc), vanadium (V), and yttrium (Y) that are much stronger than allowed by the PHOENIX grid. We find, using the spectral synthesis code Spectroscopy Made Easy, that [Sc/Fe] may be an order of magnitude above solar. For comparison, we also observed  an empirical calibrator in NGC6791, the highest metallicity cluster known ([M/H] $\sim 0.4$). Most lines are well matched between the calibrator and the Galactic center stars, except for Sc, V, and Y, which confirms that their abundances must be anomalously high in these stars. These unusual abundances, which may be a unique signature of nuclear star clusters, offer an opportunity to test models of chemical enrichment in this region.
\end{abstract}


\section{Introduction}

The metallicity of stars and stellar populations is important for understanding their formation and subsequent evolution. Stellar abundances can also serve as a signature to separate multiple populations of stars formed at different times. The comparison of stellar abundance patterns has been crucial to understanding the chemical enrichment and evolution of the Milky Way \citep[e.g.,][]{2001ApJ...554.1044C,2004ApJ...601..864T,2010A&A...522A..32R}. The emergence of large spectroscopic surveys in the past two decades has propelled this field forward with observations from a wide variety of environments in the Milky Way \citep[e.g.][]{2015ApJ...808..132H,2017AJ....153...75K}. The nuclear star cluster at the center of the Milky Way gives us a unique opportunity to study the formation and evolution of a galactic nucleus. It is the only nucleus where we can obtain resolved spectra of stars lying less than 1~pc from a supermassive black hole, making it possible to study the effects of star formation and dynamical evolution in this environment \citep[e.g.][]{2009ApJ...703.1323D,2010RvMP...82.3121G,2010ApJ...708..834B,2011ApJ...741..108P,2013ApJ...764..154D}.

\begin{figure}[hbt!]
\includegraphics[width=3.5in]{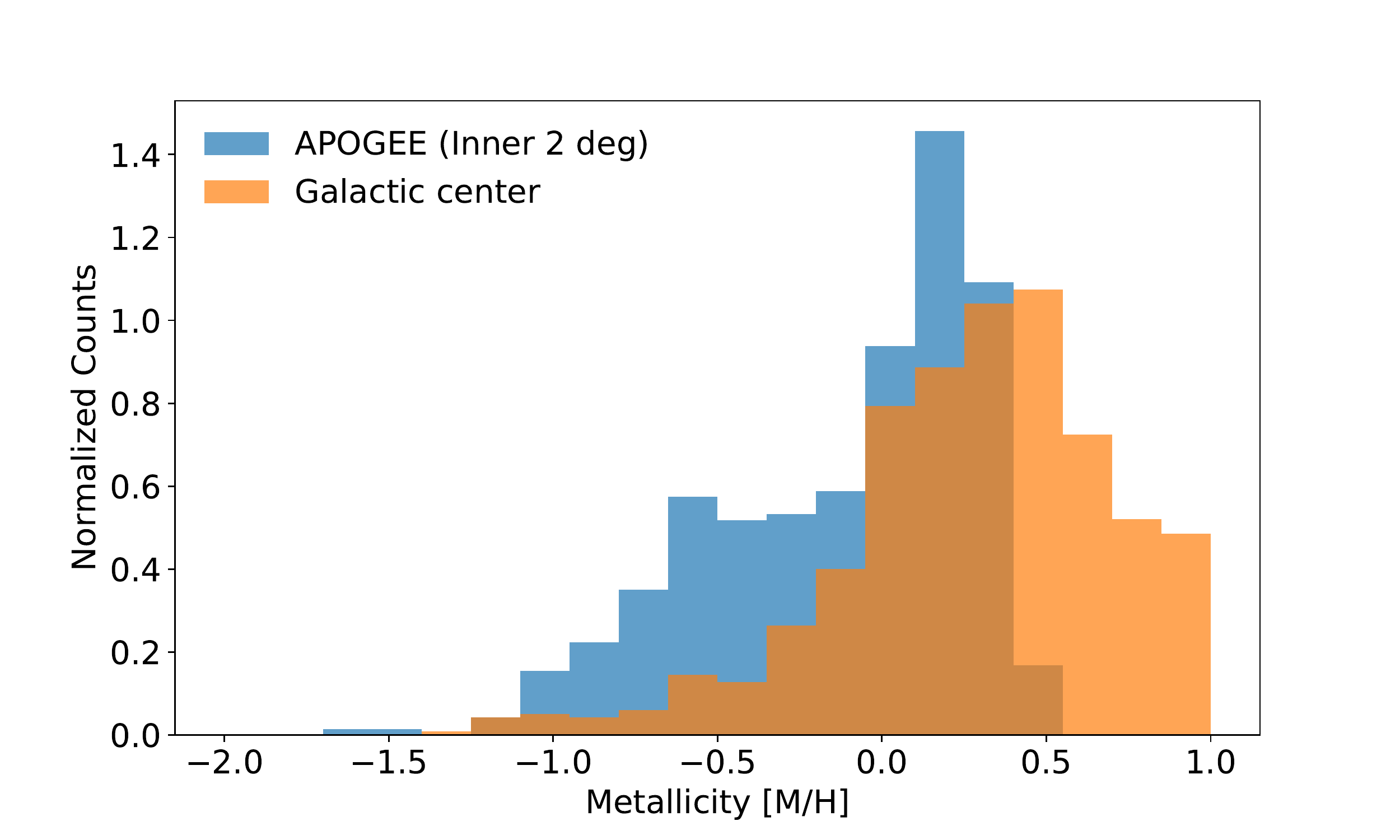}
\caption{Comparison of the metallicity distribution from the inner 2 degrees of the Milky Way from APOGEE \citep{2015AJ....150..148H} to the distribution at the Galactic center from \citet{2015ApJ...809..143D} (inner 40$^{\prime\prime}$ or 1 pc). The distribution at the very center of the Galaxy appears to be significantly offset from that of the inner bulge, suggesting a different chemical enrichment history. However, there may be systematic errors for stars at high metallicity because of model mismatch \citep{2015ApJ...809..143D,2017MNRAS.464..194F}. In order to investigate these sources, we obtain high-spectral resolution observations of two of the super-solar metallicity sources in this work.}
\label{fig:metallicity_dist}
\end{figure}

Recently, using medium-resolution spectroscopy, \citet{2015ApJ...809..143D} found that stars in the Galactic center might have even higher metallicities (Fig. \ref{fig:metallicity_dist}, [M/H] $>$ 0.5) than observed in any star clusters or in the inner bulge of the Milky Way \citep[see also,][]{2017MNRAS.464..194F}. However, these studies employed limited spectral resolution (R $\sim$ 5000) and were mismatched to the models at high metallicities. The MARCS and PHOENIX spectra grids used by \citep{2015ApJ...809..143D} and \citep{2017MNRAS.464..194F} do not appear to be able to reproduce the strength of some of the observed lines of stars at the Galactic center in K-band (1.9 to 2.4 $\mu$m). This measurement of a large variation in metallicity has the potential to change our interpretation of the star formation history, the origin, and evolution of the nuclear star cluster. However, there are two major issues unresolved by the existing observations and analyses: (1) synthetic spectra have trouble fitting observed spectra of stars at high metallicity in the K-band; (2) there has been an insufficient number of calibration studies done at K-band to improve the models. The metallicity of these stars is likely subject to large systematic uncertainties without accurate calibrations. Individual stellar abundances, which are necessary for comparison with chemical evolution models, are even less certain. Higher spectral resolution measurements are necessary to identify specific lines that are poorly fit. Observations of calibrators with accurate stellar abundance measurements at other wavelengths are necessary to disentangle model uncertainties from actual changes in abundances.

Here we report on observations at high-spectral resolution with adaptive optics of two Galactic center sources and a calibration star in the high-metallicity open cluster NGC6791. We describe the observations of these sources in Section \ref{sec:observations}. In Section \ref{sec:methodology}, we discuss the methodology used to compare model spectra to the observations, both with a model grid and spectral synthesis. In Section \ref{sec:results}, we present our resulting analyses, including identification of unusually strong Sc, V, and Y lines. In Section \ref{sec:discussion}, we discuss some implications of these results and compare them to other spectroscopic observations in the Galactic center region.

\section{Observations \& Data Reduction}
\label{sec:observations}
Observations were made using NIRSPAO, which consists of the NIRSPEC spectrograph behind laser-guide star adaptive optics, on 2016-04-29 and 2016-05-16 UT. We used the echelle mode of NIRSPEC, which results in spectral resolution of $R \sim 25000$. In order to capture the relevant wavelength regions, we used the K-band filter with the echelle cross-disperser setting of 63.85/35.8. This setting provided coverage from about 2.075 $\micron$ to 2.273 $\micron$ (orders 34, 35, 36, \& 37). These orders cover much of this wavelength range, except for some gaps between orders. The laser guide-star was propagated at the center of the field of view for each observation, and for low-order tip-tilt corrections, we used the R = 13.7 mag star, USNO 0600-28577051 (17:45:40.720 -29:00:11.20), which is located $\sim 19\arcsec$  from Sgr A*.

We observed the Galactic center stars NE1-1 001 (K = 10.4 mag) and NE1-1 002 (K = 10.7 mag) \citep{2015ApJ...808..106S,2015ApJ...809..143D}; these stars were targeted because of their unusually high metallicity inferred from spectral-template fitting \citep{2015ApJ...809..143D}. These stars are located about 0.5 pc from the black hole, Sgr A*, in projection. We also observed a comparison source in NGC6791, the highest metallicity cluster known, with [M/H] = 0.4 \citep{2007A&A...473..129C}. The source, J19213390+3750202, also has observations from APOGEE in H-band, which we will use to compare consistency between the models. All observations were made using ABBA nods to remove sky lines from consecutive nods. We observed standard A stars to correct for atmospheric absorption of telluric lines.

The data were reduced using the standard NIRSPEC reduction package, REDSPEC\footnote{\url{https://www2.keck.hawaii.edu/inst/nirspec/redspec}}. This package provides spatial rectification using standard stars as well as wavelength solutions from etalon lamp observations. We also used the spectral extraction routine in REDSPEC to obtain the final spectra.

\section{Spectral Fitting Methodology}
\label{sec:methodology}
We used two methods to fit the observed spectra: spectral template fitting with the PHOENIX grid of synthetic spectra \citep{2013A&A...553A...6H} using the Bayesian inference tool \textit{Starkit} \citep{2015zndo.soft28016K} and spectral synthesis with Spectroscopy Made Easy \citep[SME,][]{1996A&AS..118..595V,2017A&A...597A..16P}. The PHOENIX grid was used to determine the overall physical parameters such as effective temperature ($T_{eff}$), surface gravity ($\log g$), metallicity ([M/H]), and alpha-abundance ([$\alpha$/Fe]). Here, we define the $\alpha$-elements as O, Ne, Mg, Si, S, Ar, Ca, and Ti. SME was used to constrain the abundances of individual elements after fixing the global parameters from \textit{Starkit}. The methods are described below.

\subsection{Spectral template fitting with the PHOENIX Grid}

Spectral template fitting with the PHOENIX grid was performed using the same method described in \citet{2015ApJ...808..106S} with the software package \textit{StarKit}. In brief, \textit{StarKit} is a modular spectral fitting framework using Bayesian inference to determine the best-fit parameters and their uncertainties. \textit{StarKit} simultaneously fits the physical parameters of stars, the spectral continuum, the radial velocity, and the rotational velocity \citep{2015zndo.soft28016K}. The set of physical parameters available for fitting is determined by the parameters sampled by the spectral grid. For the PHOENIX grid, the parameters are $T_{eff}$, $\log g$, [M/H], and [$\alpha$/Fe]. We chose to use the PHOENIX grid for this work compared to the MARCS grid \citep{2008A&A...486..951G} in \citet{2015ApJ...808..106S} because the PHOENIX grid samples [$\alpha$/Fe] and has a more accurate treatment of the depth and shapes of spectral features. We use a linear interpolator to interpolate between the synthetic spectral grid points. We then convolve the spectral resolution of the grid to R = 25000 in order to match the instrumental resolution.

Statistically, the fit is done by computing the posterior distribution in Bayes' Theorem	:
\begin{equation}
P(\theta| D) = \frac{P(D|\theta)P(\theta)}{P(D)}
\end{equation}
where $D$ is the observed spectrum, and the model parameters $ \theta = (T_{eff}, ~\log g, ~\textrm{[M/H]}, \textrm{[}\alpha\textrm{/Fe]}, v_z,v_{rot})$, where $v_z$ is the radial velocity and $v_{rot}$ is the rotational velocity. The priors on the model parameters are $P(\theta)$ and $P(D)$ is the evidence, which acts as the normalization. The combined likelihood for an observed spectrum is:
\begin{equation}
P(D|\theta) = \prod^{\lambda_n}_{\lambda=\lambda_0}\frac{1}{\epsilon_{\lambda,obs}\sqrt{2\pi}}\exp{(-(F_{\lambda,obs} - F_\lambda(\theta))^2/2\epsilon_{\lambda,obs}^{2})},
\end{equation}
where $F_{\lambda,obs}$ is the observed spectrum, $F_{\lambda}(\theta)$ is the model spectrum evaluated with a given set of model parameters, and $\epsilon_{\lambda,obs}$ is the 1 $\sigma$ uncertainty for each observed flux point. This likelihood assumes that the uncertainty for each flux point is approximately Gaussian. For computational efficiency, we use the log-likelihood in place of the likelihood:
\begin{equation}
\ln P(D|\theta) \propto -\frac{1}{2} \sum^{\lambda_n}_{\lambda=\lambda_0}((F_{\lambda,obs} - F_\lambda(\theta))^2/\epsilon_{\lambda,obs}^{2}).
\end{equation}

We use flat priors in all the fitting parameters and sample the posterior using MultiNest, a nested sampling implementation \citep{2009MNRAS.398.1601F}. We fix the value of $\log g$ to 1.0 for all fits based on allowed $\log g$ values for their luminosity and temperature in the Parsec isochrones \citep{2012MNRAS.427..127B}. Fixing the surface gravity is useful because $\log g$ is not well constrained using the K-band spectra alone \citep{2015ApJ...808..106S,2017MNRAS.464..194F}. We use the peak posterior value to be the best fit values and the marginalized 68\% central confidence intervals for each fit parameter to be its uncertainty. Based on the tests against empirical references described in \citet{2017MNRAS.464..194F}, we include a systematic uncertainty term added in quadrature to the statistical uncertainties of each fit of $\sigma_{T_{eff}} = 200$ K, $\sigma_{[M/H]} = 0.2$, $\sigma_{[\alpha/Fe]} = 0.2$.

\subsection{Spectral Synthesis}
In order to investigate individual abundances, we use Spectroscopy Made Easy \citep[SME,][]{1996A&AS..118..595V,2017A&A...597A..16P}. This code requires a stellar atmosphere model, a line list, micro-turbulence parameters, and a solar composition to synthesize spectra. By adjusting the relative abundance of an element compared to the bulk composition, one can generate spectra to constrain the effect of elemental abundance changes on spectral features. We used the LTE MARCS12 atmospheric model with 2 km s$^{-1}$ micro-turbulence built into SME \citep{2008A&A...486..951G}, the atomic and molecular line list from VALD \citep{2015PhyS...90e4005R}, and the solar composition from \citet{2007SSRv..130..105G}. We update the VALD line list for Sc (scandium) using the newer empirical measurements of $\log gf$ values and excitation energy from \citet{2015A&A...582A..98P}. This results in a significant change in the strengths of two of the Sc I lines (22030.265 \& 22071.255 Angstrom) in the K-band when compared to PHOENIX spectra, which were computed before these measurements.

To examine the effect of a change in abundance of individual elements, we set the SME global parameters $T_{eff}$, $\log g$, [M/H], [$\alpha$/Fe] to be the best fitted values from the \textit{StarKit} fit.

\section{Results}
\label{sec:results}
\subsection{PHOENIX Grid Fit}
\label{sec:high_metallicity_compare}

The \textit{Starkit} fit to the two Galactic center stars show one star with slightly above solar metallicity, one with super-solar metallicities and both stars with about solar [$\alpha$/Fe] abundance (Table \ref{tab:results}). The posterior distribution for [M/H] for the hight metallicity star (NE1-1 002) has a distribution that peaks at the edge of the PHOENIX grid at [M/H] = 1.0, which likely indicates that we only have a lower limit on the metallicity. We, therefore, report the 95\% confidence lower limit on [M/H] for this star. NE1-1 001 and the calibration star has a well constrained posterior for [M/H] so we report the peak value as usual.
The best fit parameters for NE1-1 001 is [M/H] $ = 0.14\pm0.2$, [$\alpha$/Fe] = $-0.01\pm0.2$, $T_{eff}$= $3029\pm200$ K, while the best fit for NE1-1 002 is [M/H] $> 0.6$, [$\alpha$/Fe] = $-0.01\pm0.02$, $T_{eff}$= $3284\pm200$ K.  For comparison, the best fit for J19213390+3750202 in NGC6791 results in [M/H] = $0.18\pm0.2$, [$\alpha$/Fe] = $0.17\pm0.2$, $T_{eff}$= $3956\pm200$. These values are consistent with that from APOGEE (see Section \ref{sec:models}). If the value of $\log g$ is allowed to be free in the fit for J19213390+3750202, we obtain a best fit of [M/H] = $0.55\pm0.23$, [$\alpha$/Fe] = $0.16\pm0.21$, $T_{eff}$= $3845\pm210$, $\log g = 1.6\pm1.0$. These results are also consistent with the APOGEE results. Table \ref{tab:results} summarizes these measurements. The PHOENIX grid has difficulty in matching some of the features in the K-band spectrum, especially at super-solar metallicities (see below).

\begin{deluxetable*}{lrrrrr}
\tablecolumns{6}
\centering
\tabletypesize{\scriptsize}
\tablecaption{Results of PHOENIX grid fit}
\tablehead{\colhead{Name} &  \colhead{K} & \colhead{Teff} & \colhead{$\log g$\tablenotemark{a}} & \colhead{[M/H]} & \colhead{[$\alpha$/Fe]} }
\startdata
                 NE1-1 001 &  10.4 & $3029\pm200$ K &   1.0 &  $0.14\pm0.2$ &      $-0.01\pm0.2$ \\
                 NE1-1 002 &  10.7 & $3284\pm200$ K &   1.0 &  $> 0.6$ &      $-0.01\pm0.2$ \\
 NGC6791 J19213390+3750202 & 8.8   & $3956\pm200$ K &   1.0 &  $0.18\pm0.2$ &       $0.17\pm0.2$
\enddata
\tablenotetext{a}{The surface gravity ($\log g$) is fixed to a typical value for giants with the observed luminosity and temperature of the  stars.}
\label{tab:results}
\end{deluxetable*}

\subsection{SME Spectral Synthesis}

\begin{figure*}

\centering
\includegraphics[width=6.in]{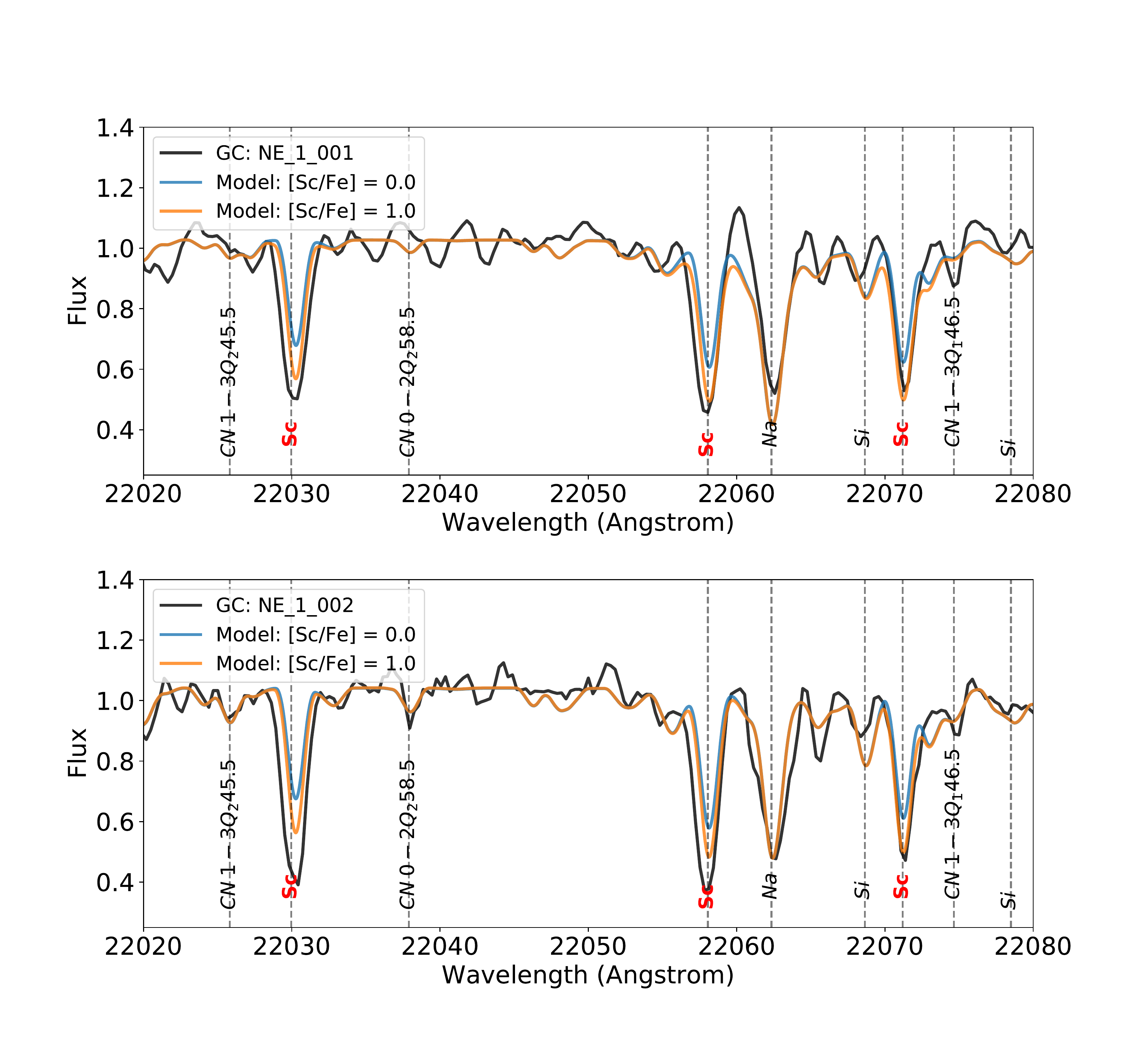}
\caption{A comparison between the observed spectra of the two Galactic center sources NE1-1 001 (top, black) and NE1-1 002 (bottom, black) and the SME synthesized spectra. The SME spectra are synthesized with the best-fit parameters determined using \textit{Starkit} while varying the [Sc/Fe] abundance. We show two examples at [Sc/Fe] =0.0 (blue) and [Sc/Fe] = 1.0 (orange). The Sc abundance seems likely to be substantially above solar. However, uncertainties are difficult to determine at this time due to lack of empirical comparisons at these high abundances.}
\label{fig:sme_fit}
\end{figure*}

The strongest mismatch between the model and the data are the Sc lines, so we have concentrated our spectral synthesis with SME to investigate these features. We use the best fit parameters for $T_{eff}$, $\log g$, [M/H], and [$\alpha$/Fe] from \textit{Starkit} while varying the [Sc/Fe] abundance with SME. For some Sc lines (22030.265 \& 22071.255 Angstrom), the spectra generated from SME show much stronger lines than from the PHOENIX model. This is likely due to the updates of the $\log gf$ and excitation energy values from \citet{2015A&A...582A..98P}. The PHOENIX grid was generated with a line list from before these measurements were made. Even with the new Sc atomic data, we find that the [Sc/Fe] values for the Galactic center sources need to be increased by over an order of magnitude with respect to solar values, or more, in order to match the observed lines in the Galactic center sources (Fig. \ref{fig:sme_fit}).

\subsection{Empirical comparisons}

\begin{figure*}[htb]
\center
\includegraphics[width=\linewidth]{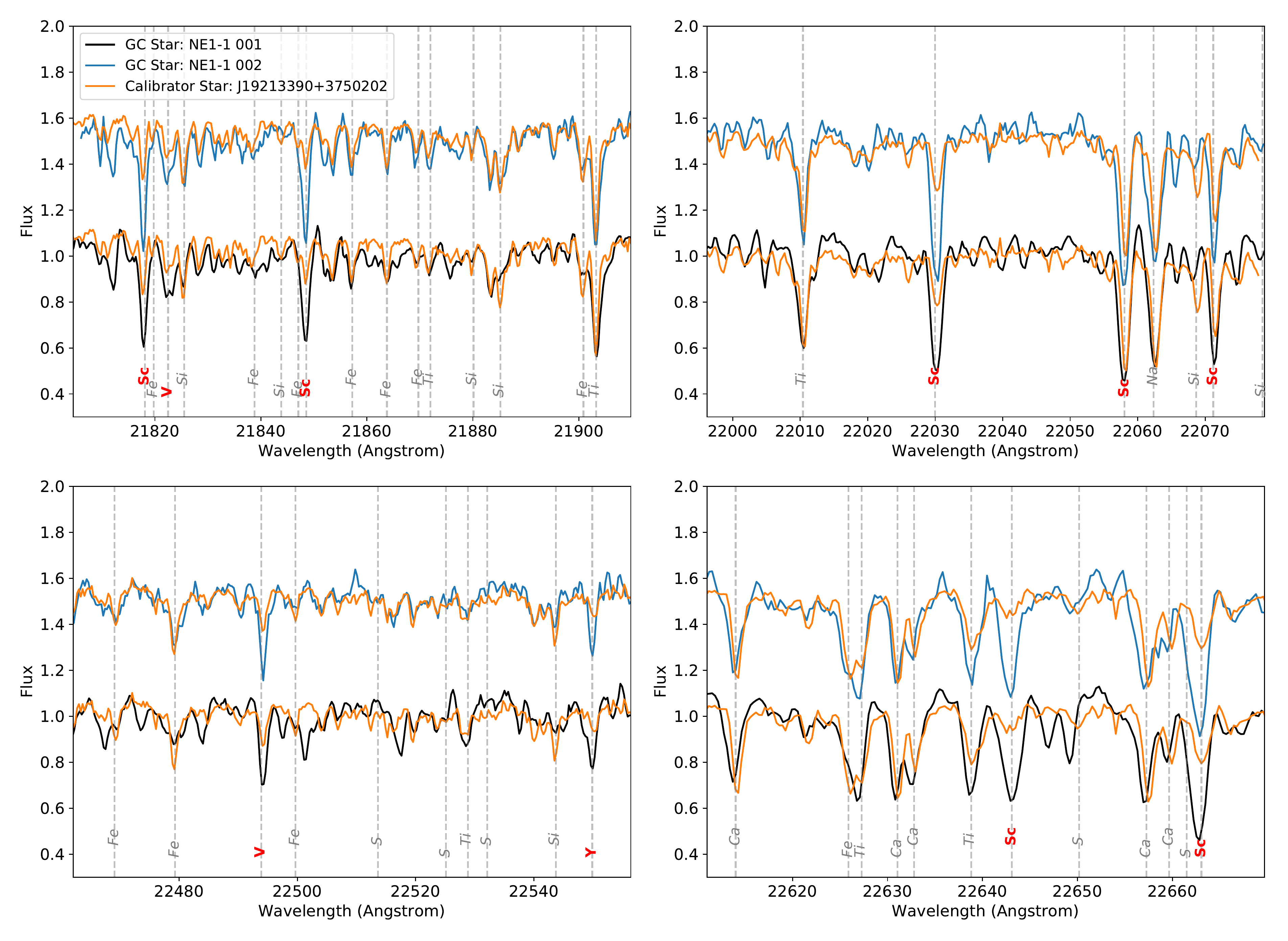}
\caption{Empirical comparison between J19213390+3750202 in NGC6791 with the Galactic center stars NE1-1 001 and NE1-1 002 in order 35 of NIRSPEC. The two panels show wavelength ranges of some of the lines of Sc, V, and Y.  The spectra are offset by 0.5 in the y-axis for clarity. Most of the spectral features are well matched except for Sc, V, and Y.}
\label{fig:ngc6791_compare35}
\end{figure*}

We use the star J19213390+3750202 in NGC6791, observed in the same way as the Galactic center sources, as an	 empirical comparison. This source was also observed as a calibrator for ASPCAP, the SDSS APOGEE abundance measurement pipeline \citep{2013AJ....146..133M}. The best fit global parameters from ASPCAP are: $T_{eff} = 3802 \pm 69$ K, $\log g = 1.33 \pm 0.08$, [M/H] = $0.3 \pm 0.02$, and [$\alpha$/Fe] = $0.14\pm 0.01$. We use this star as a qualitative comparison to the Galactic center sources, independent of the spectral model fitting, which appears to have difficulties at high metallicities. J19213390+3750202 in general matches the observed spectra of NE1-1 001 and NE1-1 002 fairly well (Figure \ref{fig:ngc6791_compare35}). The exceptions are for the Sc, V, and Y lines, which are much stronger in the Galactic center stars, with NE1-1 002 showing the strongest deviation.

\section{Discussion}
\label{sec:discussion}

\subsection{Comparison to other observations}

We have confirmed that at least one of the stars identified by \citet{2015ApJ...809..143D} as having super-solar metallicity does appear to have high metallicities using high-resolution spectroscopy. These new observations have also allowed us to identify the major source of spectral mismatch is likely from Sc, V, and Y (see below). 

These observations are also consistent with recent results from the metallicity distribution of the inner bulge. \citep{2017arXiv171201297G} showed using APOGEE spectra that there are two peaks in the metallicity distribution in the inner bulge, one at [M/H] = 0.0 and one at [M/H] = 0.32. The super-solar metallicity peak is very similar to the peak [M/H] = 0.4 in \citep{2015ApJ...809..143D} and [M/H] = 0.3 in \citep{2017MNRAS.464..194F} found in the nuclear star cluster. Recent high-spectral resolution observations in the K-band of stars in the nuclear star cluster show a metallicity distribution peak at slightly below solar [M/H] = -0.16 with a few stars at super-solar metallicities \citep{2016AJ....151....1R,2017AJ....154..239R}. These studies show a trend of increasing metallicity toward the Galactic center, but more observations will be necessary to establish whether stars within the nuclear star cluster follows this trend and whether it is distinct in abundances. 

\subsection{Models in K-band at super-solar metallicities}
\label{sec:models}

While observations of very metal-poor stars have motivated improvements in the ability of models to reproduce observed spectral features at low metallicities \citep[e.g.][]{1995AJ....109.2757M,2003ApJ...591..936S,2004A&A...416.1117C}, there have been few studies of super-solar metallicity stars beyond observations of open clusters such as NGC6791 and globular clusters such as Terzan 5 \citep{2013AJ....146..133M,2014ApJ...795...22M}. Our results show that more work will be required to theoretically model higher metallicity stars. Both the PHOENIX grid and SME synthetic spectra show mismatches. There was evidence from \citet{2015ApJ...809..143D} and \citet{2017MNRAS.464..194F} that this is the case for medium resolution spectra, but we have now confirmed that even at high-spectral resolution, the issues persist.

Since the PHOENIX grid does not go above [M/H] = 1.0, and there are no known empirical calibrators at these metallicities, it is difficult at this time to assess the accuracy of the fit for NE1-1 001 and NE1-1 002. The PHOENIX grid fit to our comparison source, J19213390+3750202 has a $[M/H] = 0.18\pm0.2$, which is consistent, though less precise than the SDSS value of $[M/H] = 0.3 \pm 0.02$. The other fitted physical parameters ($T_{eff}$ \& [$\alpha$/Fe]) for J19213390+3750202 are also consistent with their SDSS ASPCAP values. Future studies will be required of more stars having existing accurate metallicity measurements in order to determine whether there is a systematic offset in the models used here.

\subsection{Anomalous abundances in Sc, V, and Y}

While the lack of accurate models for the stars at high metallicity limits our ability to measure individual stellar abundances at this time, the comparisons to NGC6791 show that the abundances of certain elements are likley elevated in the Galactic center sources. The lines of Sc, V, and Y are especially strong in the Galactic center sources compared to the same lines in our empirical calibrator in NGC6791. The abundance of [Sc/Fe] was measured for several stars in NGC6791 by \citet{1998ApJ...502L..39P} and \citet{2007A&A...473..129C}. It ranges from about 0.0 to -0.2 depending on the star and Sc lines. Meanwhile, other metal lines such as Fe, Ti, Na, and Si are slightly stronger for the Galactic center sources compared to J19213390+3750202, but the lines of these elements in the cluster and the Galactic center are much more similar than the lines of Sc, V, and Y. This suggests that the mismatch in Sc, V, and Y is likely due to higher relative values of [Sc/Fe], [V/Fe] and [Y/Fe] rather than simply differences in overall metallicity [M/H], [$\alpha$/Fe], $T_{eff}$, or $\log g$.  Changes in these bulk physical properties of the star should change both the depth and shape of many lines, not just Sc, V, and Y. The Sc abundance might be a factor of 10 higher in the Galactic center stars than in NGC6791. Earlier observations from \citet{2000ApJ...530..307C} of the star IRS 7 (K = 6.7 mag), a red-super giant at the Galactic center, also found very strong lines of Sc and V compared to sources outside the Galactic center. \citet{2000ApJ...530..307C} suggested that perhaps these stronger lines may be attributed to differences in the atmosphere of IRS 7 rather than from increased abundances in these elements. Our work suggests that these differences are more likely to be from abundance differences as these stars are likely fairly different than IRS 7, being 4 magnitudes fainter in luminosity, and yet showing the same spectral feature anomalies. However, more work will be necessary to derive more accurate estimates of the lines. For example, investigations in non-LTE effects, multi-dimensional radiative transfer, and curve-of-growth effects would be helpful to obtain better measurements. 

\begin{figure}[hbt]
\includegraphics[width=3.25in]{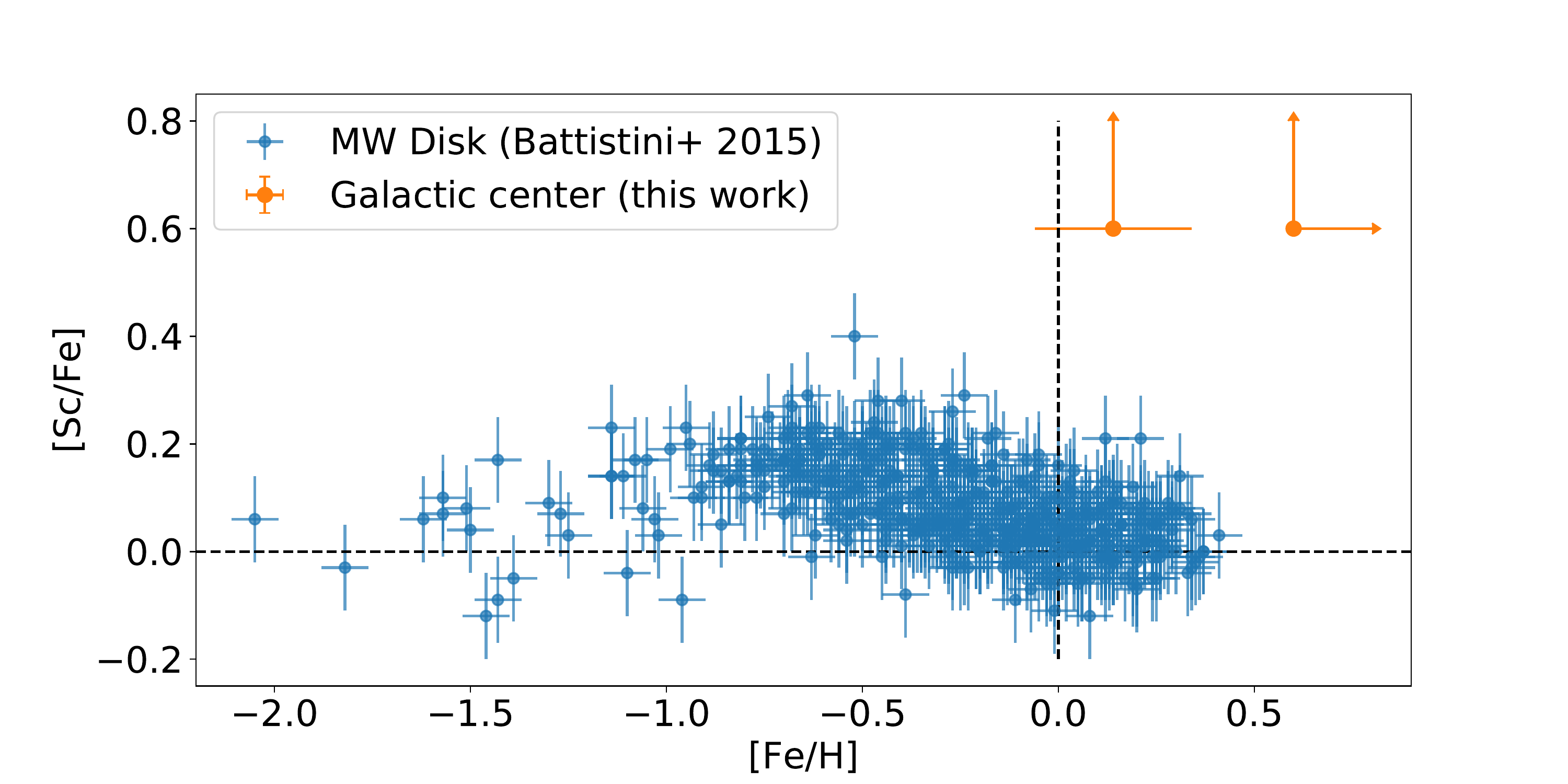}
\caption{The relationship between [Sc/Fe] and [Fe/H] observed for stars the Milky thin and thick disk \citep[blue,][]{2015A&A...577A...9B} compared to the results from this work. The orange point is the 95\% lower limit on the metallicity and [Sc/Fe] abundance from the stars in this work. The [Sc/Fe] abundances of both sources in this study fall significantly outside of Galactic trends.}
\label{fig:sc_fe}
\end{figure}

While [Sc/Fe], [V/Fe] and [Y/Fe] appear to be significantly enhanced at the Galactic center for the high-metallicity stars, it is unclear what physical mechanisms are responsible for their unusual abundance. Though both Sc and V are iron-peak, Sc is mostly produced by Type II supernovae (SNe) with a small contribution from type Ia SNe, while V is primarily produced by Type Ia SN \citep{2016A&A...593A.125S,2015A&A...577A...9B}. Y is primarily an s-process element likely produced in AGB stars \citep{2004ApJ...601..864T}. The fact that these elements are enhanced relative to Fe at high metallicity is also against trends seen elsewhere in the galaxy where the [Sc/Fe] abundances fall with increasing metallicity \citep[Figure \ref{fig:sc_fe}; e.g.,][]{2000AJ....120.2513P,2015A&A...577A...9B}. These unusual properties may indicate that the chemical enrichment history of this region is very different than any other region of the Milky Way \citep{2006ApJ...653.1145K}. Metallicity and abundance measurements for a larger sample of stars may offer a powerful way to constrain the formation of our nucleus. Anomalous abundances also offer important insights into the yields of
high energy transients \citep[such as for the high Mn abundance]{2011MNRAS.414.2709S}. Further study will be required both in terms of theoretical modeling of the spectra and empirical comparisons will be necessary to realize this potential.

\acknowledgements
We wish to thank the anonymous referee for helpful comments. We also thank the staff of the Keck Observatory, especially Randy Campbell, Greg Doppmann, Heather Hershey, Jason McIlroy, Gary Puniwai, Terry Stickel, and Hien Tran for all their help in obtaining the new observations. Support for this work at UCLA was provided by NSF grant AST-1412615 and the Levine-Leichtman Family Foundation. Support for W. Kerzendorf was provided by the ESO Fellowship. The W. M. Keck Observatory is operated as a scientific partner- ship among the California Institute of Technology, the University of California, and the National Aeronautics and Space Administration. The authors wish to recognize that the summit of Maunakea has always held a very significant cultural role for the indigenous Hawaiian community. We are most fortunate to have the opportunity to observe from this mountain. The Observatory was made possible by the generous financial support of the W. M. Keck Foundation.

\software{Numpy \citep{5725236}, Scipy \citep{5725236}, Astropy \citep{2013A&A...558A..33A}, Matplotlib \citep{Hunter:2007}, pymultinest \citep{2014A&A...564A.125B}, StarKit \citep{2015zndo.soft28016K}}
          
\bibliography{/u/tdo/Documents/bibtex/prelim}

\end{document}